\documentclass[aip,jcp,amsmath,amssymb,longbibliography,floatfix,reprint]{revtex4-2}

\usepackage{graphicx}
\usepackage{dcolumn}
\usepackage{bm}

\usepackage[utf8]{inputenc}
\usepackage[T1]{fontenc}
\usepackage{mathptmx}
\usepackage{etoolbox}
\usepackage{siunitx}
\usepackage{chemformula}
\usepackage{braket}
\usepackage{hyperref}

\newcommand{\oper}[1]{\hat{#1}}

\renewcommand{\H}{\oper{{H}}}

\makeatletter
\def\@email#1#2{%
 \endgroup
 \patchcmd{\titleblock@produce}
  {\frontmatter@RRAPformat}
  {\frontmatter@RRAPformat{\produce@RRAP{*#1\href{mailto:#2}{#2}}}\frontmatter@RRAPformat}
  {}{}
}%
\makeatother
\begin{document}

\title{Path integral Monte Carlo in a discrete variable representation with Gibbs sampling: dipolar planar rotor chain}
\author{Wenxue Zhang}
\author{Muhammad Shaeer Moeed}
\author{Andrew Bright}
\author{Tobias Serwatka}
\author{Estevao De Oliveira}
\author{Pierre-Nicholas Roy}%
 \email{pnroy@uwaterloo.ca.}
\affiliation{ 
Department of Chemistry, University of Waterloo, Waterloo, Ontario N2L 3G1, Canada
}%


\date{\today}

\begin{abstract}
In this work, we propose a Path Integral Monte Carlo (PIMC) approach based on discretized continuous degrees of freedom and rejection-free Gibbs sampling.
The ground state properties of a chain of planar rotors with dipole-dipole interactions are used to illustrate the approach. Energetic and structural properties are computed and compared to exact diagonalization and Numerical Matrix Multiplication for $N\leq 3$ to assess the systematic Trotter factorization error convergence. For larger chains with up to $N=100$ rotors, Density Matrix Renormalization Group (DMRG) calculations are used as a benchmark. We show that using Gibbs sampling is advantageous compared to traditional Metroplolis-Hastings rejection importance sampling. Indeed, Gibbs sampling leads to lower variance and correlation in the computed observables.
\end{abstract}

\maketitle

\section{Introduction}
The simulation of correlated many-body quantum systems is both of great relevance and also a challenging task.
Of particular interest is the case of confined dipolar molecules, where intermolecular interactions can lead to rich quantum phases.
For example, water molecules trapped in beryl\cite{gorshunov2016incipient,kolesnikov2016quantum,zhukova2019quantum,belyanchikov2020dielectric} or cordierite\cite{belyanchikov2022single} crystals exhibit intriguing features such as ferroelectricity, quantum tunneling, and collective excitations.
Ferroelectricity has also been observed in water inside carbon nanotubes.\cite{luo2008ferroelectric} Moreover, a quasi-phase transition has been reported for the specific case of water trapped in (6, 5) carbon nanotubes.\cite{ma2017quasiphase}

Simulations play a key role in our understanding of these systems, and the inclusion of quantum effects is of prime importance. The treatment of rotational degrees of freedom is required in order to model the above systems. It is possible to describe the many-body states of correlated dipolar rotors using exact diagonalization (ED) techniques when a few molecules are present.\cite{felker2017accurate,halverson2018quantifying} 
For large systems, the exponential growth of the Hilbert space makes ED calculations impractical. For one-dimensional chain systems, tensor network techniques have successfully been used to compute the ground state of collections of dipolar linear rotors,\cite{iouchtchenko2018ground,mainali2021comparison} chains of rotating water molecules\cite{serwatka2022ground} with ferroelectric\cite{serwatka2022ferroelectric} and quantum critical behavior,\cite{serwatka2023qpt,serwatka2023endo} as well as entangled dipolar planar rotor chains.\cite{serwatka2024quantum,serwatka2024ground}
Beyond one-dimensional chains, optimizing tensor networks of Matrix Product State (MPS) form\cite{schollwock2011density} using techniques such as the Density Matrix Renormalization Group (DMRG) method\cite{white1992density} becomes prohibitively expensive.

In order to simulate higher dimensional systems such as 2-d molecular lattices or 3-d crystals of confined molecular rotors, alternate approaches are needed.
The Path Integral Monte Carlo (PIMC) method\cite{barker1979,ceperley1995path} is a very powerful tool that is amenable to the simulation of general systems in arbitrary dimensions. The generalization of PIMC to molecular rotations is well established\cite{marx1999path} and has been used to simulate the ground state of correlated dipolar linear rotors in 1-d\cite{abolins2011ground,abolins2013erratum,sahoo2020path,sahoo2023effect} and 2-d\cite{abolins2018quantum} lattices, and chains of rotating water molecules.\cite{sahoo2021path}
Path integral simulations of molecular systems are usually performed in the continuous position representation.
As an alternative, Xiao and Poirier have suggested the use of a Discrete Variable Representation (DVR)\cite{light1985generalized,light2000discrete} for PIMC simulations.\cite{xiao2007using,xiao2007efficient} 
Makri and co-workers have also used a DVR for real-time path integral calculations.\cite{makri1992improved,topaler1993quasi,topaler1993system,topaler1994quantum,makri1995numerical,makri2018communication,makri2018modular} 
When using a DVR, the path {\em integral} becomes a path {\em sum}.

Here, we propose a PIMC approach for the simulation of dipolar rotor systems based on rejection free Gibbs sampling.\cite{4767596,doi:10.1080/01621459.1990.10476213} 
The approach is tested on linear chains of planar rotors of varying sizes and interaction strengths to showcase the suitability of the approach to simulate both ordered and disordered phases, as well as quantum critical regions.
The remainder of this paper is organized as follows: in Sec. \ref{methods}, we present methodological details such as the form of the Hamiltonian, the path integral ground state (PIGS) formulation for rotors, the DVR representation, and the Gibbs sampling technique. We present and discuss results in Sec. \ref{results}, followed by concluding remarks in Sec. \ref{conclusions}.

\section{Methodology\label{methods}}

\subsection{Dipolar planar Rotor Chain}
\begin{figure}
\centering
\includegraphics[width=\columnwidth]{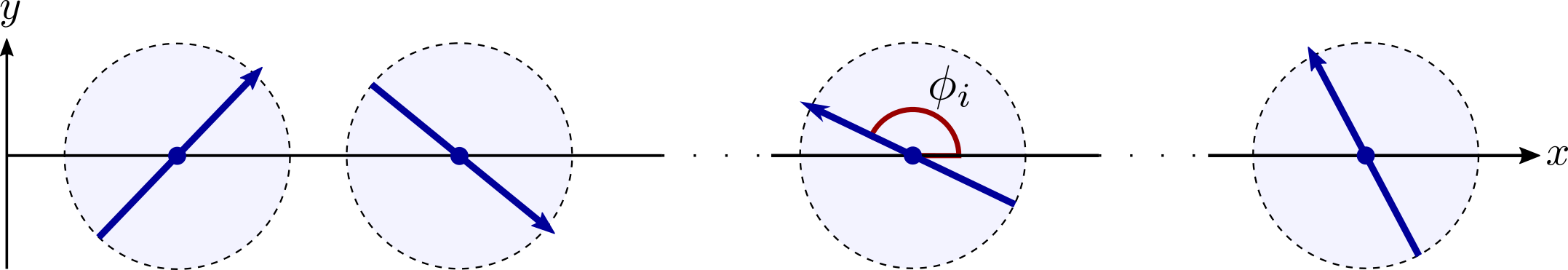}
\caption{Schematic representation of a chain of equally spaced planar rotors.\label{fig:planar_rotor_chain}	}
\end{figure}

We consider a linear chain of $N$ identical planar rotors with nearest neighbors dipole-dipole interactions. For this model, the rotations are constrained to the $xy$-plane and the linear chain has coplanar orientation along the $x$-axis as depicted in Fig. \ref{fig:planar_rotor_chain}. The Hamiltonian for this system is,
\begin{equation}
\label{eq:system_hamiltonian}
    \H=\sum_{i=1}^{N} \oper{L}_{z,i}^2+g\sum^{N-1}_{i=1}\left[\hat{y}_i \hat{y}_{i+1} - 2\hat{x}_i \hat{x}_{i+1}\right] \, ,
\end{equation}
The first term denotes the rotational kinetic energy, with $\oper{L}_{z,i}^2$ being the angular momentum corresponding to the $i$-th site. The second term represents the local dipole-dipole interaction of strength $g$ between rotors, and $\hat{x}_{i}$, $\hat{y}_{i}$ are the Cartesian components of a unit vector $\hat{e}(\phi_i)$ parametrized by the angle $\phi_i$ of the $i$-th dipole. We use units where $\hbar=1$.
In the angular position basis, the Hamiltonian becomes,
\begin{equation}
\label{eq:system_hamiltonian_angular_position_basis}
    \H = \hat{T} + g \hat{V} = -\sum_{i=1}^{N} \frac{\partial^2}{\partial \phi_{i}^2} + g\sum_{i=1}^{N-1} \hat{V}_{i,i+1}
\end{equation}
where the potential between neighboring rotors is given by, 
\begin{equation} \label{2_body_potential}
    \hat{V}_{i, i+1} = \sin(\hat{\phi}_i) \sin(\hat{\phi}_{i+1}) - 2 \cos(\hat{\phi}_i) \cos(\hat{\phi}_{i+1})
    \, .
\end{equation}
Here, $\oper{\phi}_i$ represents angular position operators acting on the continuous basis set $\{\ket{\phi_i}\}$, for $\phi_i \in \{0,2\pi\}$. It is important to notice that a further investigation of the system properties involves the explicit calculation of the matrix elements of the Hamiltonian of Eq. \eqref{eq:system_hamiltonian_angular_position_basis}. 

\subsection{Path Integrals For Rotors}
Here, we briefly review the PIGS\cite{jcp-113-1366-2000} framework to set the stage and fix our path integral notation. To this end, consider the thermal density matrix for our many-body system given by $e^{-\beta \H}$,
where $\beta=1/k_BT$ is the reciperocal temperature. We can compute the ground state expectation value of an arbitrary operator $\hat{O}$ using a trial wavefunction $|\psi_T\rangle$ and taking the limit as $\beta \rightarrow \infty$ as follows 
\begin{equation} \label{PIGS}
    \langle \hat{O} \rangle = \lim_{\beta \rightarrow \infty} \frac{\langle \psi_T | e^{-\beta \H/2} \hat{O} e^{-\beta \H/2}| \psi_T \rangle}{Z_0(\beta)}
\end{equation}
where $Z_0(\beta)$ is the ground state pseudo-partition function given by 
\begin{equation}
    Z_0(\beta) = \text{Tr}\left(e^{-\beta \H}|\psi_T\rangle \langle \psi_T| \right)
\end{equation}
If the trial state has non-zero overlap with the ground state of the system, the estimator defined by Eq. \eqref{PIGS} converges to the ground state expectation value of $\hat{O}$. 

For simplicity, we first construct our path integral using the partition function and then discuss the ground state energy PIGS estimator in a separate section. 
To this end, consider the trace evaluated in the angular position basis $\{\left|\phi_{i}\right>\}_{1 \leq i \leq N}$, 
\begin{equation} \label{partition_coords}
    Z(\beta) = \int d \bm{{\phi}} \ \rho(\bm{\phi}, \bm{\phi}, \beta)
\end{equation}
where $(S^{1})^{N}$ is the direct product of $N$ circles $(S^1)$ representing the $N$-body configuration space, $\bm{\phi} = (\phi_1, \phi_2, ..., \phi_N)$ is an arbitrary point in the $N$-body configuration space, and $\rho(\bm{\phi}, \bm{\phi}, \beta)$ represents the diagonal elements of the density matrix in the position basis,
\begin{equation}
    \rho(\bm{\phi}, \bm{\phi}, \beta) = \left<\phi_{1}, \phi_2, ... , \phi_N\right| e^{-\beta \H} \left|\phi_1, \phi_2, ..., \phi_N\right>
\end{equation}
To compute the above matrix elements of the density operator exactly, we would need to exponentiate the Hamiltonian. This is intractable for large systems due to the exponential scaling of the Hilbert space dimension with the number of rotors in the chain. However, note that we can express the low temperature density operator in terms of high temperature density operators using the convolution property, 
\begin{equation}
    e^{-\beta \H} = (e^{-(\beta/P) \H})^{P} = (e^{-\tau \H})^{P}
\end{equation}
where $P$ is some positive integer (representing the number of time-slices), and $\tau = \beta/P$ is the imaginary time. Then, inserting $P-1$ angular position basis resolutions of identity yields for the density operator matrix elements, 
\begin{equation} \label{time_slicing}
    \rho(\bm{\phi^}{1}, \bm{\phi}^{P+1}, \beta) = \int d \vec{\bm{{\phi}}} \ \prod_{i=1}^{P} \rho(\bm{\phi}^{i}, \bm{\phi}^{i+1}, \tau)
\end{equation}
where $\vec{\bm{{\phi}}} = (\bm{\phi}^1, ..., \bm{\phi}^{P+1})$.
Note that superscripts in our notation are reserved for bead numbers whereas subscripts will be used to represent particle numbers.
We can now approximate the high-temperature density using the symmetrized 
Trotter
formula which gives 
\begin{equation} \label{Trotter_abstract}
    e^{-\tau \H} = e^{-\tau\hat{V}/2}e^{-\tau \hat{T}}e^{-\tau \hat{V}/2} + \mathcal{O}(\tau^3)
\end{equation}
Here, $\hat{T}$ is the kinetic energy operator and $\hat{V}$ is the potential energy operator. This particular symmetric splitting allows us to use the fact that the potential is diagonal in the position basis to express the density matrix as follows, 
\begin{equation} \label{Trotter_coords}
    \rho(\bm{\phi}^{i}, \bm{\phi}^{i+1}, \tau) \approx \rho_{T}^{i, i+1} (\tau) \ \rho_{V}^{i} (\tau/2) \ \rho_{V}^{i+1} (\tau/2)
\end{equation}
where 
\begin{gather} \label{rho_k, rho_v}
    \rho_{V}^{i} (\tau/2) = \rho_V(\bm{\phi}^{i}, \tau/2) = e^{-\tau V(\bm{\phi}^i)/2} \\
    \rho_{T}^{i, i+1} (\tau) = \rho_T(\bm{\phi}^{i}, \bm{\phi}^{i+1}, \tau) = \left<\bm{\phi}^{i}\right| e^{-\tau \hat{T}} \left|\bm{\phi}^{i+1}\right>
\end{gather}
Using Eqs. $\eqref{partition_coords}, \eqref{time_slicing}$ and $\eqref{Trotter_coords}$, we finally get for the finite temperature partition function,
\begin{equation} \label{partition_pi}
    Z = \lim_{P \rightarrow \infty} \int d \vec{\bm{\phi}} \ \prod_{i=1}^{P} \rho_{V}^{i} (\tau/2) \ \rho_{V}^{i+1} (\tau/2) \ \rho_{T}^{i, i+1} (\tau)
\end{equation}
where $\vec{\bm{{\phi}}} = (\bm{\phi}^1, ..., \bm{\phi}^P, \bm{\phi}^{P+1})$. Here, we need to impose the constraint that the first and last bead be the same ($\bm{\phi}^1=\bm{\phi}^{P+1}$) since we are evaluating the trace. Moreover, we need to take the limit as $P \rightarrow \infty \iff \tau \rightarrow 0$ since the Trotter factorization in equation \eqref{Trotter_abstract} is only exact in this limit. Note that $Z$ is still a function of $\beta$ although we have omitted this dependence in Eq. \eqref{partition_pi} for notational brevity.

We can further factor the kinetic energy (free) propagator since the kinetic energy operator is a sum of commuting operators $\hat{T}_{j} = \hat{L}_{z,j}$ where $j$ is the index corresponding to the $j$th rotor in the chain. This yields, 
\begin{equation}
    \rho_T^{i,i+1}(\tau) = \prod_{j=1}^{N} \langle \phi_{j}^{i}| e^{-\tau \hat{T}_{j}} | \phi_{j}^{i+1} \rangle
\end{equation}
Noting that the total potential is a sum over pair-wise terms, we can similarly factor the potential energy propagator, 
\begin{equation}
    \rho_{V}^{i} (\tau/2) = \prod_{j=1}^{N-1} e^{-\tau V(\phi_{j}^i, \phi_{j+1}^{i})/2}
\end{equation}
where $V$ in the exponent here denotes the $2$-body potential as defined by Eq. \eqref{2_body_potential}. This gives for Eq. \eqref{partition_pi}, 
\begin{multline} \label{Z_pi_factored}
    Z = \lim_{P \rightarrow \infty} \int d \vec{\bm{\phi}} \ \prod_{i=1}^{P} \prod_{j=1}^{N-1} \rho_{T} (\phi^i_{j}, \phi^{i+1}_{j}, \tau) \\ \rho_{V}(\phi^{i}_{j}, \phi^{i}_{j+1}, \tau/2) \ \rho_{V} (\phi^{i+1}_{j}, \phi^{i+1}_{j+1}, \tau/2)
\end{multline}
This final set of factorizations shows that our path integral density is a product of two-variable functions. Consequently, our planar rotor chain system is especially amenable to Markov Chain Monte Carlo (MCMC) methods since the two-variable densities can be easily pre-computed on a grid and stored prior to the simulation. Moreover, as we will show below,  
the density enters our MCMC algorithms as a quotient. 
Consequently, for sequential algorithms such as Gibbs sampling that rely on single variable moves, only those terms in the full product that depend on the variable being perturbed need to be computed for each move. This greatly reduces the density calculation overhead for each step.

\subsection{PIGS Estimators}
Having developed our path integral formalism, we can now derive the PIGS ground state energy estimator used for our analysis. To this end, consider again Eq. \eqref{PIGS} with $\hat{O} = \H$. Since the density operator commutes with the Hamiltonian, we get, 
\begin{equation}
    E_{0} = \langle \H \rangle = \lim_{\beta \rightarrow \infty} \frac{\langle \psi_T | e^{-\beta \H} \H| \psi_T \rangle}{Z_0(\beta)}
\end{equation}
Inserting angular position basis resolutions of identity as before, we get for the energy estimator, 
\begin{equation}
    E_{0}  = \frac{1}{Z_0} \int d \vec{\bm{{\phi}}} \ \psi_T^*(\bm{\phi}^1) \rho(\bm{\phi}^1, \bm{\phi}^2, \beta) \H \psi_{T}(\bm{\phi}^2) 
\end{equation}
Here, $\vec{\bm{{\phi}}} = (\bm{\phi}^1, \bm{\phi}^2)$. For our simulations, we can pick $\psi_{T}(\bm{\phi}) = 1/\sqrt{2\pi}$ to be our trial wavefunction. Then, since the kinetic energy operator is a differential operator, we get, 
\begin{equation}
    E_{0}  = \frac{1}{2\pi Z_0} \int d \vec{\bm{{\phi}}} \ \rho(\bm{\phi}^1, \bm{\phi}^2, \beta) V(\bm{\phi}^2)
\end{equation}
Because the energy estimator depends on the density operator matrix elements in the position basis, we can again use the convolution property and the Trotter expansion to get, 
\begin{multline} \label{E_0_est}
    E_{0} (\beta, P) =
    \frac{1}{2\pi Z_0(\beta)} \int d \vec{\bm{\phi}} \ \prod_{i=1}^{P} \prod_{j=1}^{N-1} \rho_{T} (\phi^i_{j}, \phi^{i+1}_{j}, \tau) \\ \rho_{V}(\phi^{i}_{j}, \phi^{i}_{j+1}, \tau/2) \ \rho_{V} (\phi^{i+1}_{j}, \phi^{i+1}_{j+1}, \tau/2) \ V(\bm{\phi}^{P+1})
\end{multline}
where $\vec{\bm{\phi}} = (\bm{\phi}^1, ..., \bm{\phi}^P, \bm{\phi}^{P+1})$ as before. The error in this estimator is $\mathcal{O}(\tau^2)$ arising, as with the partition function, from the Trotter expansion. In the limit of $P \rightarrow \infty \iff \tau=\beta/P \rightarrow 0$, the estimator converges to the ground state energy. To evaluate the limit computationally, we therefore construct an increasing sequence of time slices, $\mathcal{P} = \{P_{k}\}$, and compute the above integral for each of those values of $P_{k}$. Then, we fit the sequence of energy estimates $\mathcal{E} = \{E_{P}\}_{P \in \mathcal{P}}$ using a quadratic model and extrapolate to estimate the ground state energy. 

For structural properties (functions of position only), the operator $\hat{O}$ doesn't commute with the propagator. Consequently, we must factorize Eq. \eqref{PIGS} directly. In this work, we also consider the orientational correlation of the system defined as, 
\begin{equation} \label{corr_def}
    \hat{C} = \sum_{i=1}^{N-1} \hat{e}_{i}\cdot \hat{e}_{i+1} = \sum_{i=1}^{N-1} \cos(\hat{\phi}_i - \hat{\phi}_{i+1})
\end{equation}
where $e_i = (\cos( \phi_i ), \sin( \phi_{i} ))$ is the electric dipole corresponding to the $i$th rotor in the chain. Setting $\hat{O} = \hat{C}$ in Eq. \eqref{PIGS} and expanding in the position basis gives, 
\begin{gather}
    \langle \hat{C} \rangle  = \frac{1}{Z_0} \int d \vec{\bm{\phi}} \ \psi_T^*(\bm{\phi}^1) \rho(\bm{\phi}^1, \bm{\phi}^2, \beta/2) \nonumber \\ C(\bm{\phi}^2, \bm{\phi}^3) \rho(\bm{\phi}^3, \bm{\phi}^4, \beta/2) \psi_T (\bm{\phi}^4) 
\end{gather}
where $\vec{\bm{\phi}} = (\bm{\phi}^1, \bm{\phi}^2, \bm{\phi}^3, \bm{\phi}^4)$. Since the correlation is diagonal in the position basis, we get,
\begin{equation}
    C(\bm{\phi}^2, \bm{\phi}^3) = \langle \bm{\phi}^2|\hat{C} |  \bm{\phi}^3 \rangle = \delta(\bm{\phi}^2 - \bm{\phi}^3) C(\bm{\phi}^2) 
\end{equation}
where $C(\bm{\phi}^2)$ is defined by Eq. \eqref{corr_def}. We can now evaluate the integral over $\bm{\phi}^2$ using the delta function and re-label the integration variables $(1,2,3) \rightarrow (1, P/2 + 1, P+1)$ for convenience. This yields 
\begin{gather}
    \langle \hat{C} \rangle = \frac{1}{2\pi Z_0} \int d \vec{\bm{{\phi}}} \ \rho(\bm{\phi}^1, \bm{\phi}^{P/2 + 1}, \beta/2) \nonumber \\ \rho(\bm{\phi}^{P/2+1}, \bm{\phi}^{P+1}, \beta/2) C(\bm{\phi}^{P/2 + 1})
\end{gather}
where we have used our trial function definition $\psi_{T}(\bm{\phi}) = 1/\sqrt{2\pi}$. Finally, Trotter expanding both densities in the above equation $P/2$ times each gives the PIGS estimator for the correlation,
\begin{gather}
    C (\beta, P) =
    \frac{1}{2\pi Z_0(\beta)} \int d \vec{\bm{\phi}}\ \prod_{i=1}^{P} \prod_{j=1}^{N-1} \rho_{T} (\phi^i_{j}, \phi^{i+1}_{j}, \tau) \nonumber \\ \rho_{V}(\phi^{i}_{j}, \phi^{i}_{j+1}, \tau/2) \ \rho_{V} (\phi^{i+1}_{j}, \phi^{i+1}_{j+1}, \tau/2) \ C(\bm{\phi}^{P/2 + 1})
\end{gather}
It is instructive to note here that $P$ has to be even since we assumed that $P/2$ is an integer during the time-slicing derivation above. Moreover, the density in the correlation estimator is exactly the same as that for energy and so the same MCMC simulation can be used to estimate both. However, in the case of correlation (or any structural properties) the corresponding position kernel must be evaluated at the middle bead only. This is a distinct feature of PIGS in contrast to finite temperature PIMC where all the beads are equivalent and structural properties can be evaluated on any of the beads due to the existence of closed paths.

\subsection{Grid Representation: from Path Integrals to Path Sums}
\label{sec:grid_representation_DVR}

Our proposed approach for computing ground state properties relies on discretizing the continuous angular position basis using the Colbert-Miller DVR.\cite{colbert1992novel} In this framework, the discretized angular position basis takes the form,
\begin{equation} \label{DVR_phi}
    \{\left|\phi_\alpha\right>\} = \{\left|\phi(\alpha)\right>\} = \{\left|2 \pi \alpha / L\right>\} = \{\left|\alpha\right>\}
\end{equation} 
where $L = 2m+1$ is the total number of grid points, $d$ is the Fourier representation cutoff and $\alpha$ is an integer such that $0 \leq \alpha \leq 2d$. Note that the index $\alpha$ here represents a point on the $\phi$ grid as opposed to a particle or bead number. Greek subscripts on $\phi$ will henceforth be reserved for grid indices. We will continue our convention of using Latin subscripts for particle indices. 
The $1$-body kinetic energy matrix elements in the DVR\cite{colbert1992novel} representation take the form,
\begin{multline} \label{DVR_T}
    \langle\alpha|T|\alpha'\rangle =
    \begin{cases} 
      \frac{d(d+1)}{3} & \alpha = \alpha' \\
      (-1)^{\Delta \alpha} \frac{\cos(\pi (\Delta \alpha)/L)}{2\sin^2(\pi (\Delta \alpha)/L)} & \alpha \neq \alpha' \\
   \end{cases}
\end{multline}
where $\Delta \alpha = \alpha - \alpha'$. In the DVR basis, we can diagonalize the $1$-body kinetic energy operator numerically and compute its exponential. This allows us to pre-compute and store the $1$-body kinetic energy propagator as a matrix. The $2$-body potential energy operator is evaluated on the grid as follows,
\begin{equation}
    \langle\alpha_1, \alpha_2|V|\alpha_1', \alpha_2'\rangle = V(\phi(\alpha_1), \phi(\alpha_2)) \delta_{\alpha_1, \alpha_1'}\delta_{\alpha_2, \alpha_2'}
\end{equation}
The DVR basis also enables us to store the diagonal entries of the ($2$-body) potential energy propagator as a matrix. The grid representation maps the $N(P+1)$-dimensional integrals in Eqs. \eqref{Z_pi_factored} and \eqref{E_0_est} to summations. Therefore, in this framework, the energy estimator yields,
\begin{multline} \label{path_sum}
    E_{0}(\beta, P) = \frac{1}{2\pi Z} \sum_{\vec{\bm{\alpha}}} \prod_{i=1}^{P}\prod_{j=1}^{N-1} \rho_{T} (\alpha^{i}_{j}, \alpha^{i+1}_{j}, \tau) \\ \rho_{V}(\alpha^{i}_{j}, \alpha^{i}_{j+1}, \tau/2) \ \rho_{V} (\alpha^{i+1}_{j}, \alpha^{i+1}_{j+1}, \tau/2) \ V(\bm{\alpha}^{P+1})
\end{multline}
where $\vec{\bm{\alpha}}$ is an $N(P+1)$ dimensional vector. Each entry of $\vec{\bm{\alpha}}$ represents one variable in the path sum and can take integer values between $0$ and $2d$. In this "vectorized" form, the $p$-th bead and $n$-th rotor orientation is specified by the component with index $k = (p-1)N + n$. $\bm{\alpha}^{P+1}$ is the sub-tuple of $\vec{\bm{\alpha}}$ corresponding to the last bead. Note that we have employed the following abuse of notation in specifying Eq. \eqref{path_sum},
\begin{align}
    \rho_{V}(\alpha^{i}_{j}, \alpha^{i}_{j+1}, \tau/2) = \rho_{V}(\phi(\alpha^{i}_{j}), \phi(\alpha^{i}_{j+1}), \tau/2) \\
    \rho_{T}(\alpha^{i}_{j}, \alpha^{i+1}_{j}, \tau) = \rho_{T}(\phi(\alpha^{i}_{j}), \phi(\alpha^{i+1}_{j}), \tau) \\
    V(\bm{\alpha}^{P+1}) = V(\phi(\alpha_{1}^{P+1}), ..., \phi(\alpha_{N}^{P+1}))
\end{align}
As discussed above, one benefit of the DVR grid is that it enables us to pre-compute all three of these components $(\rho_T, \rho_V, V)$ in the discretized angular position basis (Eq. \eqref{DVR_phi}) and index them as required during the simulation. Not only does this discretization optimize the efficiency of the MCMC simulation but also enables us to use sampling approaches adapted to discrete probability distribution such as Gibbs sampling.

\subsection{Gibbs Sampling The Many Body Distribution}

MCMC approaches approximate multi-dimensional integrals (such as equation \eqref{partition_pi}) stochastically, 
\begin{equation}
    \int_{\Omega} d\bm{x} \ f(\bm{x}) p(\bm{x}) \approx \frac{1}{M} \sum_{\bm{x}_i \in S} f(\bm{x}_{i})
\end{equation}
where $\Omega$ is the multi-dimensional configuration space, $p(\bm{x})$ is some probability distribution, and $S$ is a set of $M$ points in $\Omega$ statistically sampled according to $p(\bm{x})$. We can also use this approach to estimate multi-dimensional sums such as Eq. \eqref{path_sum}. For the rotor chain energy estimator in the DVR basis, the probability distribution is
\begin{multline} \label{full_prob}
    p(\vec{\bm{\alpha}}) = \frac{1}{2\pi Z} \prod_{i=1}^{P} \prod_{j=1}^{N-1} \rho_{T} (\alpha^{i}_{j}, \alpha^{i+1}_{j}, \tau) \\ \rho_{V}(\alpha^{i}_{j}, \alpha^{i}_{j+1}, \tau/2) \ \rho_{V} (\alpha^{i+1}_{j}, \alpha^{i+1}_{j+1}, \tau/2)
\end{multline}
Owing to the exponential scaling of the number of variables with the number of particles, this $N$-body distribution is difficult to sample directly. However, we can leverage the factorized form of the density by instead sampling a Markov chain that converges to the desired distribution. 

Conventional approaches in this domain involve using the sequential Metropolis-Hastings (M-H) algorithm in which, at each step, a move is proposed and then either accepted or rejected depending on the likelihood of the proposed configuration compared to the previous configuration. This invariably leads to a set of samples with a high degree of redundancy because a large set of proposed moves are rejected. In this work, we adapt the sequential Gibbs sampling algorithm to the planar rotor problem which, by construction, is rejection free and consequently, more efficient than M-H. The algorithm, in steps, is described below:
\begin{enumerate}
    \item Specify the initial configuration of all the variables:
    \begin{equation}
        \left(\vec{\bm{\alpha}}\right)^0 =\left(\alpha^0_1, ..., \alpha^0_{(P+1)N}\right)
    \end{equation} 
    \item To Sample $(\vec{\bm{\alpha}})^{k+1}$, given a configuration for the current step $(\vec{\bm{\alpha}})^k$, loop over all the $(P+1)N$ components, sampling each sequentially. To sample $\alpha^{k+1}_{j}$, we use the conditional distribution, 
    \begin{align}
    \label{eq:conditional_prob}
        p\left(\tilde{\alpha}^{k+1}_{j} | \alpha^{k+1}_{1}, .., \alpha^{k+1}_{j-1},\alpha^{k}_{j+1}, .., \alpha^{k}_{(P+1)N}\right) = \nonumber \\
        \frac{1}{Z_G}p\left(\alpha^{k+1}_{1}, .., \alpha^{k+1}_{j-1},\tilde{\alpha}^{k+1}_{j},\alpha^{k}_{j+1}, .., \alpha^{k}_{(P+1)N}\right)
    \end{align}
    where $\tilde{\alpha}_{j}^{k+1}$ are indices ${1,...,N-1}$ specifying angular position values on the DVR grid and $Z_G$ is a normalization. The sampled value of $\tilde{\alpha}^{k+1}_{j}$ is used as a parameter to sample $\tilde{\alpha}^{k+1}_{j+1}$.
    \item Repeat step 2 for each pass until the desired statistical accuracy is obtained. Finally, compute the average potential energy over the sampled configurations to estimate the ground state energy.
\end{enumerate}

Note that we have reserved superscripts to denote the sample index and subscripts to denote components of the sample. Therefore, $\alpha^i_j$ denotes the $j$th component of the $i$th sample.
As shown in Eq. \eqref{eq:conditional_prob}, the full probability density only enters the algorithm as a quotient. Therefore, the factorized form of the density (Eq. \eqref{full_prob}) simplifies the single-variable sampling greatly,
    \begin{gather}
        p\left(\tilde{\alpha}^{k+1}_{j} | \alpha^{k+1}_{1}, .., \alpha^{k+1}_{j-1},\alpha^{k}_{j+1}, .., \alpha^{k}_{(P+1)N}\right) = \nonumber \\
        p\left(\tilde{\alpha}^{k+1}_{j_{p,n}}|\alpha^{k}_{j_{(p+1),n}},\alpha^{k+1}_{j_{(p-1),n}},\alpha^{k}_{j_{p,(n+1)}},\alpha^{k}_{j_{p,(n-1)}}\right)
    \end{gather}
where the subindex $j_{p,n}$ represents now the "vectorized" index associated with bead $p$ and rotor $n$ given by $j_{p,n} = (p-1)N + n$. Since there is a finite set of values that $\tilde{\alpha}^{k+1}_{j_{p,n}}$ can take, we can sample this conditional via weighted uniform sampling.

It is also instructive to note that in the conditional distribution for the $j$th variable, samples from step $(k+1)$ are used for the first $j-1$ variables and samples from step $k$ are used for the last $N(P+1) - j$ variables. Therefore, the conditional distribution changes for each variable within each step. This implies that the associated Markov chain is not reversible and the sampler does not satisfy detailed balance (similar to the sequential M-H technique). However, the algorithm does satisfy the balance condition which implies that the associated Markov chain still converges to the desired probability distribution.\cite{Manousiouthakis_1999,Faizi_2020}

\section{Results and Discussion\label{results}}

To assess the utility of the DVR grid as a viable discretization scheme for Gibbs sampling, we first study the imaginary time convergence of the density for small numbers of particles ($N = 2$ and $N = 3$). Then, we extend the approach to larger system sizes ($N = 100$) and study their structural properties. In particular, we find that using the Gibbs-DVR approach results in faster Markov chain convergence due to lower correlations between steps. Moreover, this methodology is also simpler to optimize than rejection-based techniques such as continuous M-H because it doesn't have a tunable acceptance ratio modulating the efficiency of the simulation. 

We studied the convergence of ground state energy for $N=2$ to determine the minimum required frequency cutoff for ground state properties using the DVR grid. We find that $m = 5$ ($L=11$ grid points for each rotor) is sufficient. However, as discussed below, for any given value of $m$, there exists a value of $\tau$ such that the density acquires a sign problem due to the presence of positive off-diagonal entries in the Hamiltonian in the position basis. This renders the density unusable for sampling techniques. This issue can be remedied if necessary by increasing the value of $m$. For our results, $m = 5$ is sufficient since we do not require $\tau < 0.2$ for any of our results to see convergence. The approach discribed in Ref. \onlinecite{serwatka2024ground} is used for the DMRG benchmark calculations. Calculations are performed using the iTensor package.\cite{fishman2022itensor}

\subsection{Imaginary Time Step Convergence}
\begin{figure}[t]
  \centering
\includegraphics[width=\columnwidth]{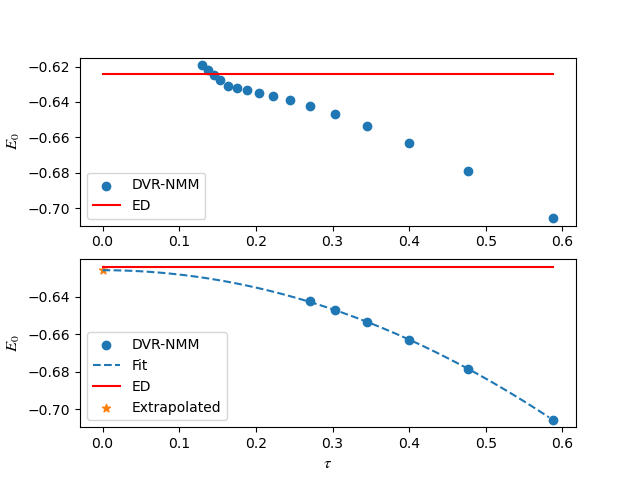}\caption{Convergence of the DVR-NMM $E_0$  with $\tau$ compared to ED results with maximum number of DVR grid points $L = 11$, $\beta = 10.0$ and $g = 0.6$ for $N = 2$ (top panel). Fit of the $\tau > 0.17$ results and extrapolation to $\tau =0$ (bottom panel).\label{figEvstauN2} } 
\end{figure}

\begin{figure}[h]
  \centering
\includegraphics[width=\columnwidth]{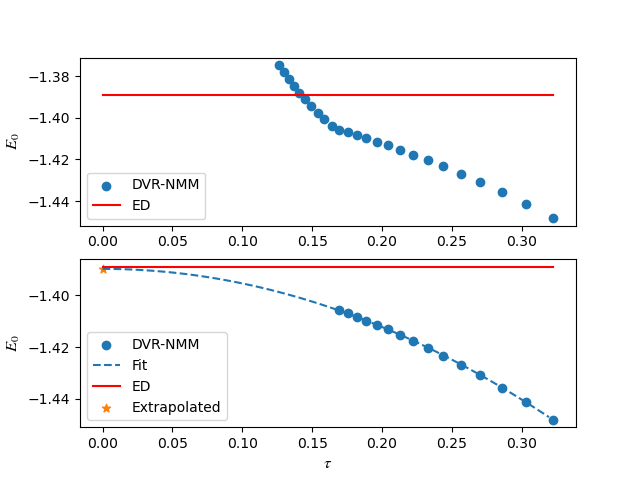}
\caption{Same as Fig. \ref{figEvstauN2} for $N=3$ here.\label{figEvstauN3} } 
\end{figure}

To test convergence with $\tau$, we compute the ground state energy using the DVR grid and the Numerical Matrix Multiplication (NMM) method\cite{nmm_ref} for $N = 2$ and $N = 3$ and compare with ED results. The purpose of this NMM analysis is to highlight the convergence behavior of the systematic factorization error.
As shown in Figs. \ref{figEvstauN2} and \ref{figEvstauN3}, fitting NMM results corresponding to $\tau > 0.17$ results in convergence to ED results. However, below a critical $\tau$ value the NMM results diverge. This issue has been raised by Xiao and Poirier.\cite{xiao2007using,xiao2007efficient}

We found that this divergent behavior at small $\tau$ values comes from the $1$-body kinetic energy only and therefore does not depend on system size $N$ or  interaction strength $g$.
The DVR representation of the kinetic energy (see Eq. \eqref{DVR_T}) contains positive entries on the off-diagonals due to the cosine function. This results in the DVR representation of the Hamiltonian having non-negative off-diagonal entries for $m > 1$. Consequently, the Hamiltonian in the DVR basis is non-stoquastic. For small $\tau$, $e^{-\tau H} \approx {1} - \tau H$ which means that the off-diagonal entries of the density matrix are given by: $\langle \phi | e^{-\tau H} | \phi' \rangle \approx -\tau \langle \phi | T | \phi' \rangle$. Therefore, these matrix entries are negative if $\langle \phi | T | \phi' \rangle > 0$. Consequently, we can't interpret the DVR basis representation of the density as a set of transition probabilities and use them in sampling paradigms such as PIMC when $\tau$ is small for this problem. Note that this sign problem doesn't affect NMM. The divergence shown in Figs. \ref{figEvstauN2} and \ref{figEvstauN3} is due to the fact that we use the absolute values of the propagator matrix entries since PIMC requires a positive-definite density. Finally, it should also be noted that this critical value of $\tau$ can be made smaller by increasing $m$ in the DVR representation. 

\subsection{Reciprocal Temperature Convergence}
\begin{figure}[h]
    \includegraphics[width=\columnwidth]{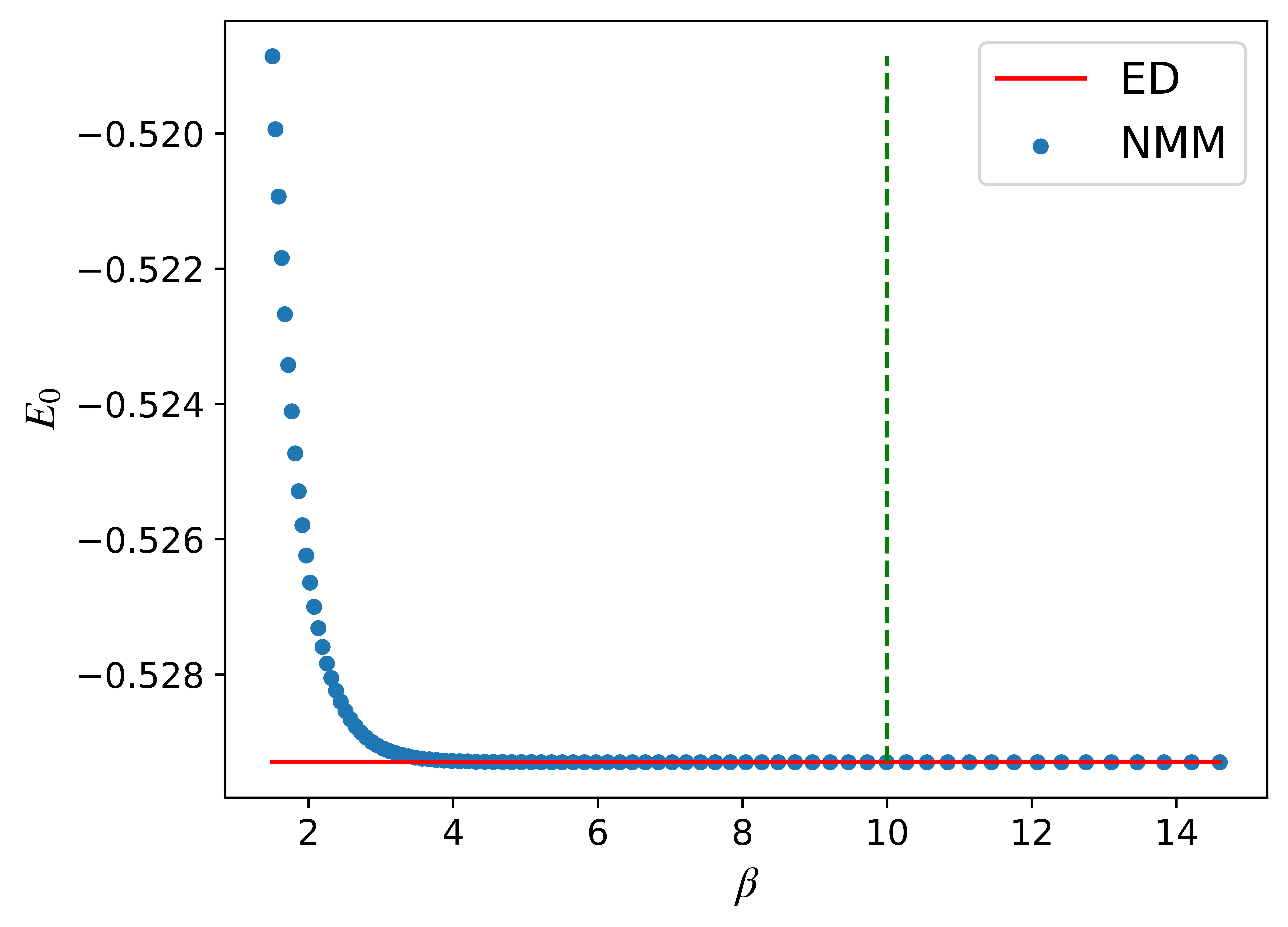}
    \caption{Convergence of the DVR-NMM $E_0$ with $\beta$ compared to ED  for $N = 2$ and $g = 1.0$. The NMM results were computed by fitting a sequence of seven DVR-NMM estimates corresponding to $\tau_{t} = 0.18 + 0.01 t$ for $0 \leq t \leq 7$, $t \in \mathbb{Z}$ using: $E_\tau = E_0 + a \tau^2 + b \tau^4$. $\beta = 4.0$ appears to be sufficient for convergence. We use $\beta = 10.0$ for our analysis (highlighted in the figure).}
    \label{fig:betaconvergence}
\end{figure}

As discussed above, our path integral simulations converge to the ground state operator expectation values in the limit of $\beta \rightarrow \infty$. Therefore, we study the convergence of the PIGS estimator as a function of $\beta$ to determine a finite value $\beta_c$ such that the associated error for all $\beta \geq \beta_c$ is negligible. To this end, we computed ground state energy estimates $E_0(\tau, \beta)$ using DVR-NMM for $N = 2$ corresponding to a decreasing sequence of $\tau$ values: $\{\tau\}_{k} = \{0.24 - 0.01k\}$ for integers $0 \leq k \leq 7$. Fitting a quadratic model in $\tau$ with an additional quartic term ($E(\tau, \beta) = E_0(\beta) + a(\beta) \tau^2 + b(\beta) \tau^4$) for each $\beta$, we computed the $\tau$-independent DVR-NMM estimates of the ground state energy associated with logarithmically spaced values of $\beta$ in the interval $(1.5,15)$ as shown in Fig. \ref{fig:betaconvergence}. We find that at $\beta \approx 4.0$, the error is negligible which is not surprising given the exponential dependence of the propagator on $\beta$. Since any value above $4.0$ is sufficient, we pick $\beta = 10.0$ for our $N$-body simulations. 

\begin{figure}[h]
\centering
\includegraphics[width=\columnwidth]{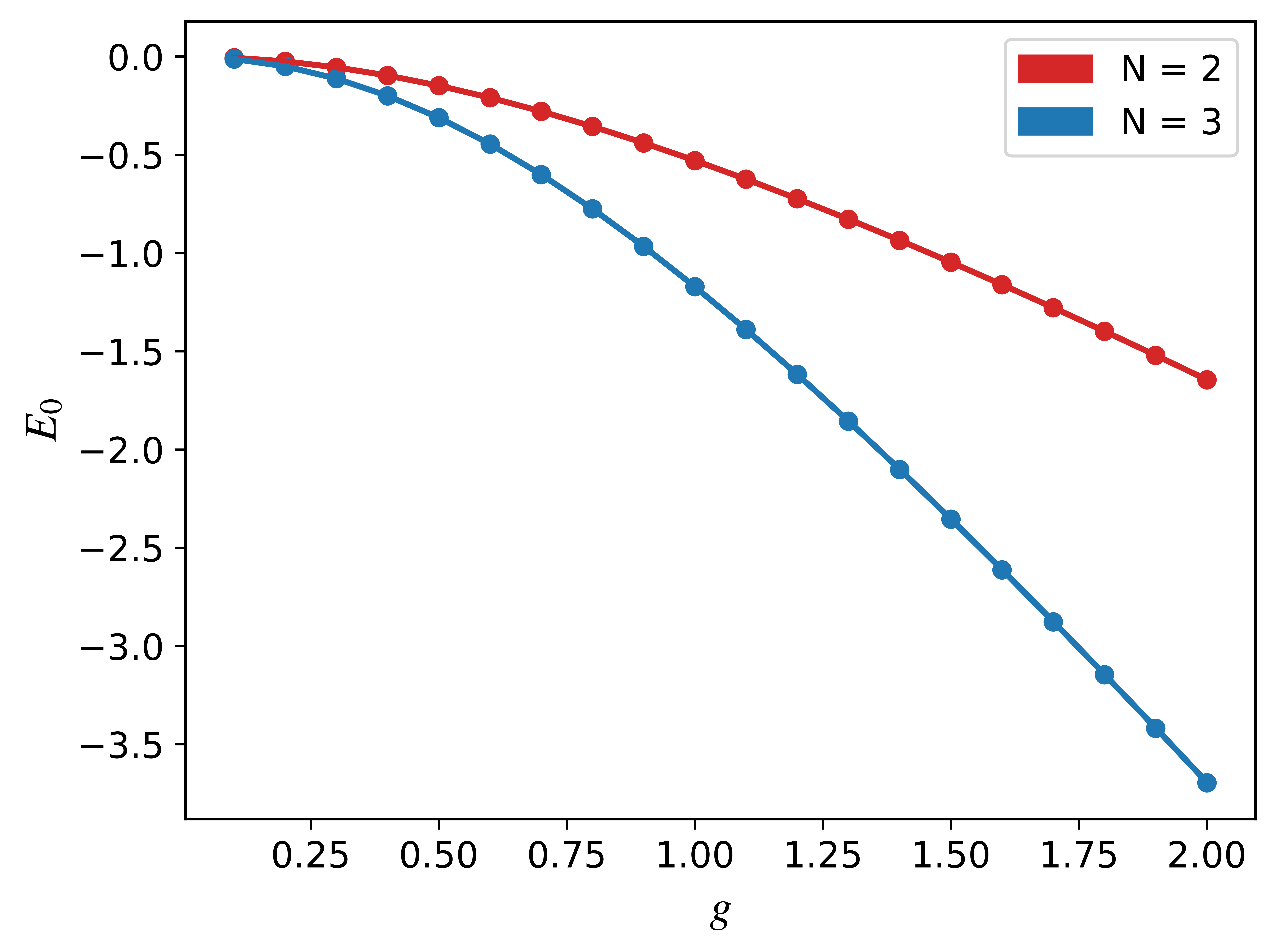}
    \caption{Ground state energy as a function of the coupling strength $g$ computed using NMM for $N = 2$ and $N = 3$. The solid lines indicate the ED results while the points specify the NMM estimates (evaluated by extrapolating $E_0$ NMM estimates for small $\tau$ to $0$). In both cases, the $\tau$-extrapolated DVR-NMM results converge to the ED results for all $g \in (0.1, 2.0)$.}
\label{fig:small_N_energy}
\end{figure}

Finally, we also compute the ground state energies as a function of the coupling strength $g$ for $N = 2$ and $N = 3$ by extrapolating in $\tau$ as discussed above with $\beta = 10.0$. Comparing this to ED results (see Fig. \ref{fig:small_N_energy}), we see that the error is negligible. This suggests that our methodology should be robust to variations in the interaction strength $g$ and the number of rotors in the chain $N$. We further examine this point in the next section while using our approach to study this system in the large $N$ limit. 

\subsection{Many-Body Extension}

To showcase the utility of this approach for studying the ground state properties of large systems, we compute the ground state energy and the orientational correlation function. Since we cannot diagonalize the Hamiltonian for large systems, we compare our results to DMRG instead. Figure \ref{PIGS_DMRG_E0} below shows the ground state energy for $N = 100$ and Fig \ref{PIGS_DMRG_corr} shows the correlation for $N = 100$. The solid lines show the DMRG results. For both the energy and the correlation, the PIGS estimates converge to the DMRG results within statistical error (estimated using the binning method). For both of these properties, we use $\beta = 10.0$, $P = 49$ and $50000$ steps for the PIGS estimates. The simulations for $g < 0.5$ were started with randomly oriented rotors whereas for $g \geq 0.5$, the simulations were started with all the rotors pointing along the chain since we expect all the rotors to be aligned at equilibrium when $g$ is large. 

\begin{figure}
  \centering
\includegraphics[width=\columnwidth]{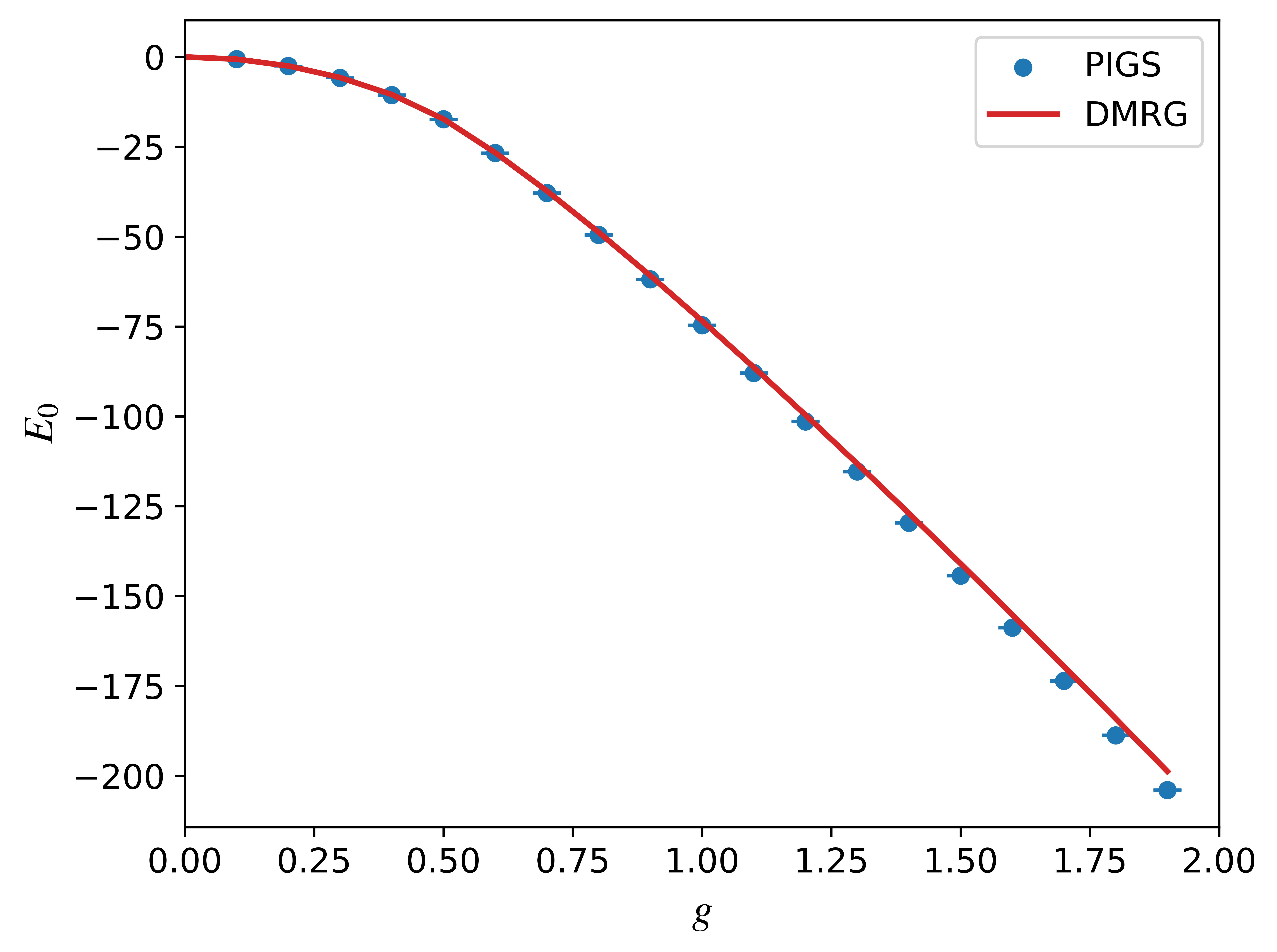}
\caption{Ground state energy for $N = 100$, $P = 49$ computed using PIGS with DVR-Gibbs sampling. DMRG results plotted for comparison.
} \label{PIGS_DMRG_E0}
\end{figure}

\begin{figure}
  \centering
\includegraphics[width=\columnwidth]{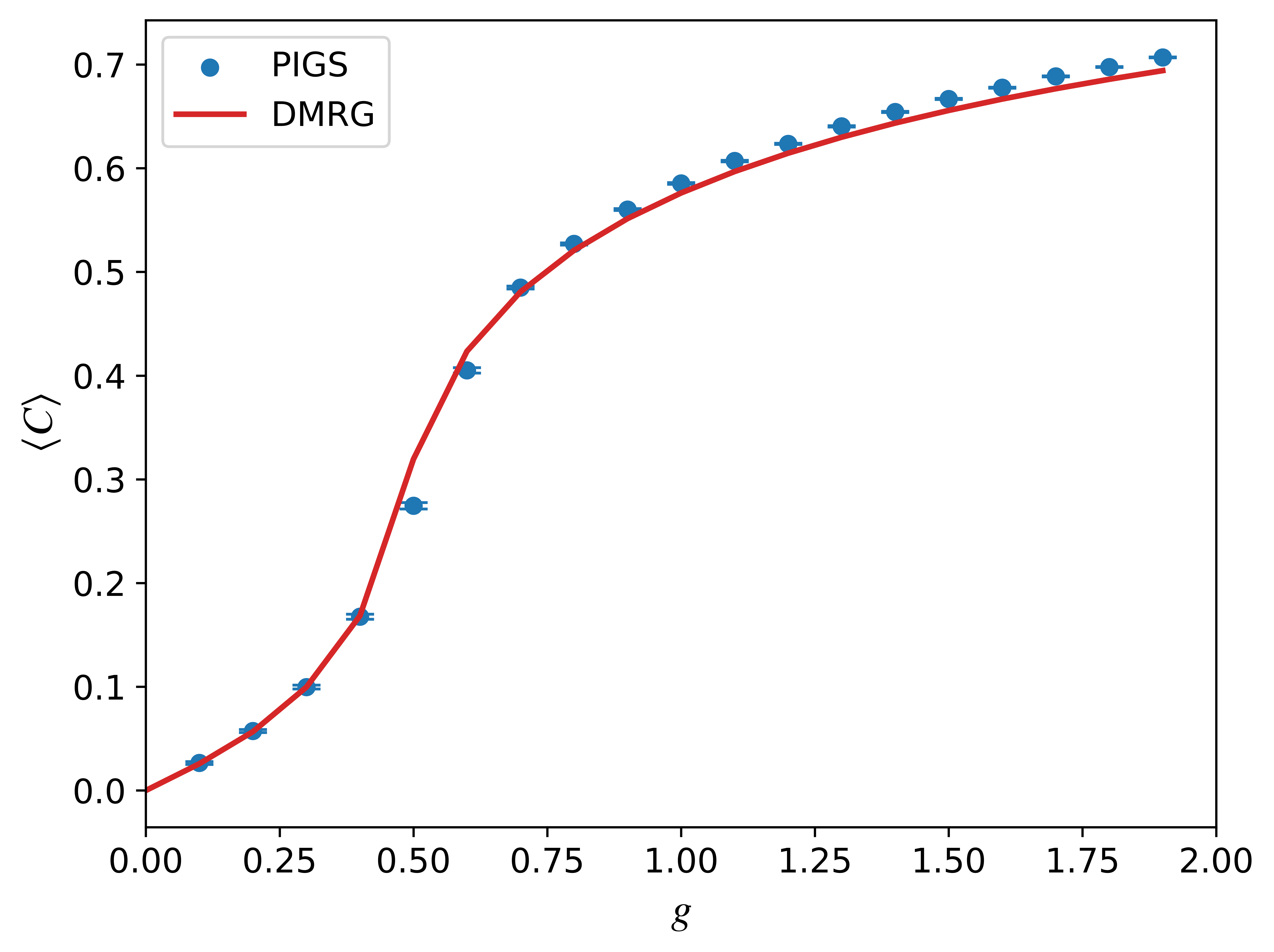}
\caption{Correlation for $N = 100$, $P = 49$ computed using PIGS with DVR-Gibbs sampling. DMRG results are plotted for comparison.
} \label{PIGS_DMRG_corr}
\end{figure}

Previous work has shown that there is a phase transition in this system at $g \sim 0.5$ as the ground state moves from the disordered state ($g < 0.5$) to an ordered state ($g > 0.5$).\cite{serwatka2024quantum} While typically, phase transitions are difficult to study using Monte Carlo simulations due to the critical slow down near the transition, the existence of these two phases in the large $N$ limit for our system can be resolved by our approach as can be seen from Fig. $\ref{PIGS_DMRG_corr}$. 


\begin{figure}
    \centering
    \includegraphics[width=\columnwidth]{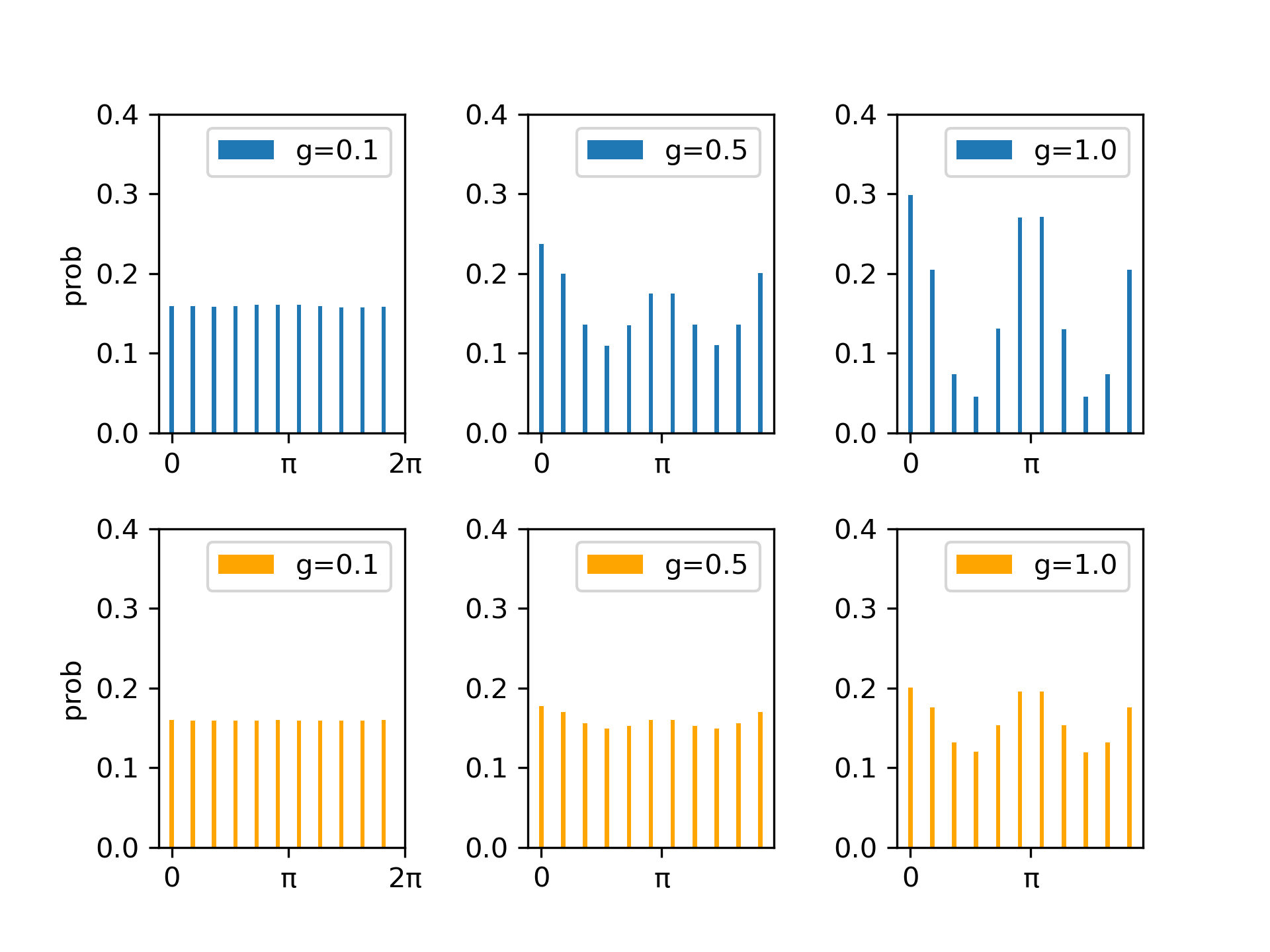}
    \caption{Angular distributions of the middlde beads (top panel) and end beads (bottom panel) for $N=100$ with interactions strengths $g=0.1, 0.5, 1.0$.}
    \label{fig:histograms}
\end{figure}


Angular distributions for $N=100$ are shown in Fig. \ref{fig:histograms}. We selected three interaction strength values ($g=0.1, 0.5, 1.0$).
The distributions are uniform when $g=0.1$, since $g$ is less than the critical value ($g_c=0.5$).
At $g=0.5$ some localization starts to appear and the chain is polarized at $\pi$ and $2\pi$ when $g=1.0$.

\begin{figure*}[t]
    \includegraphics[width=.33\textwidth]{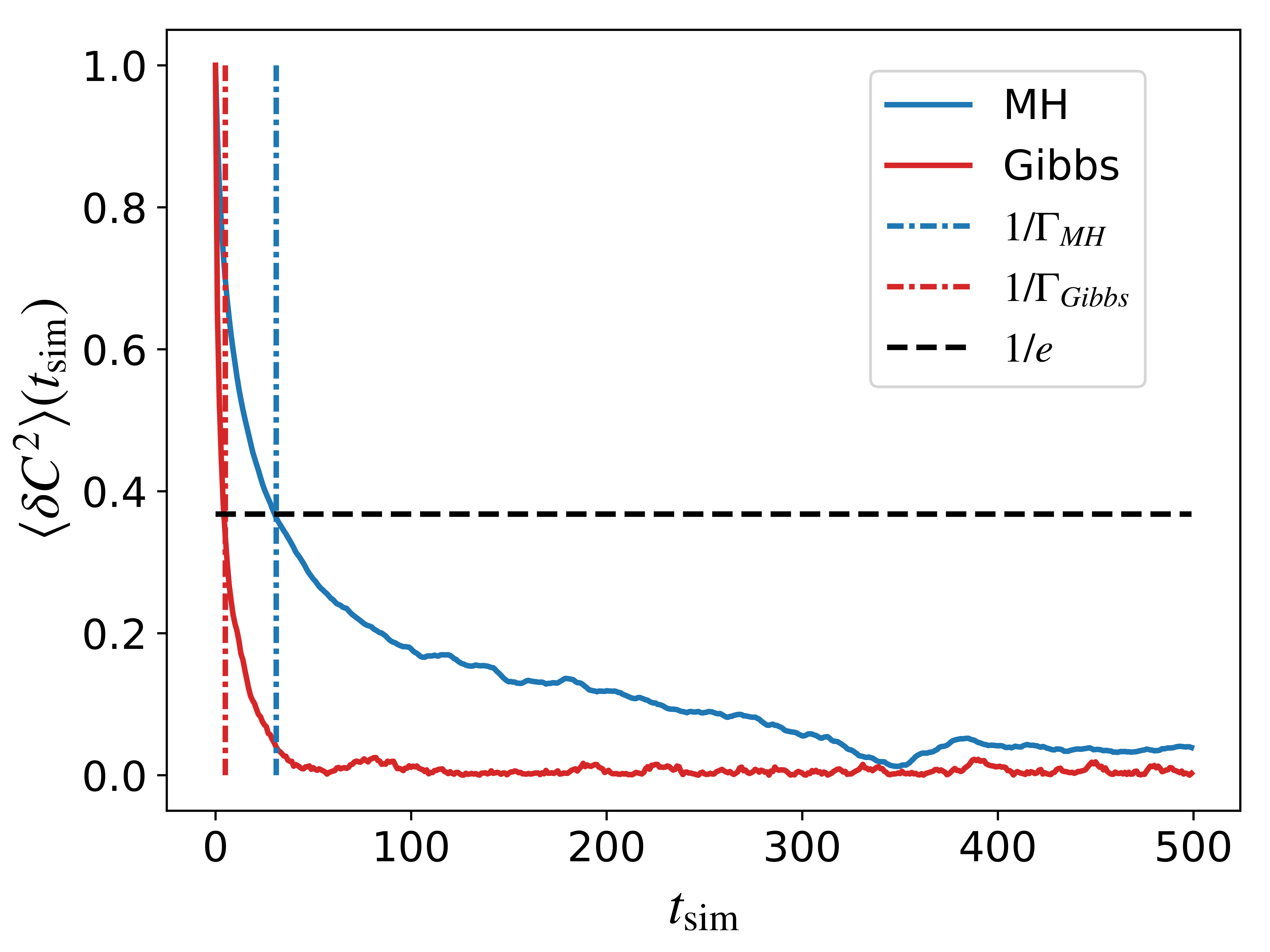}
    \includegraphics[width=.33\textwidth]{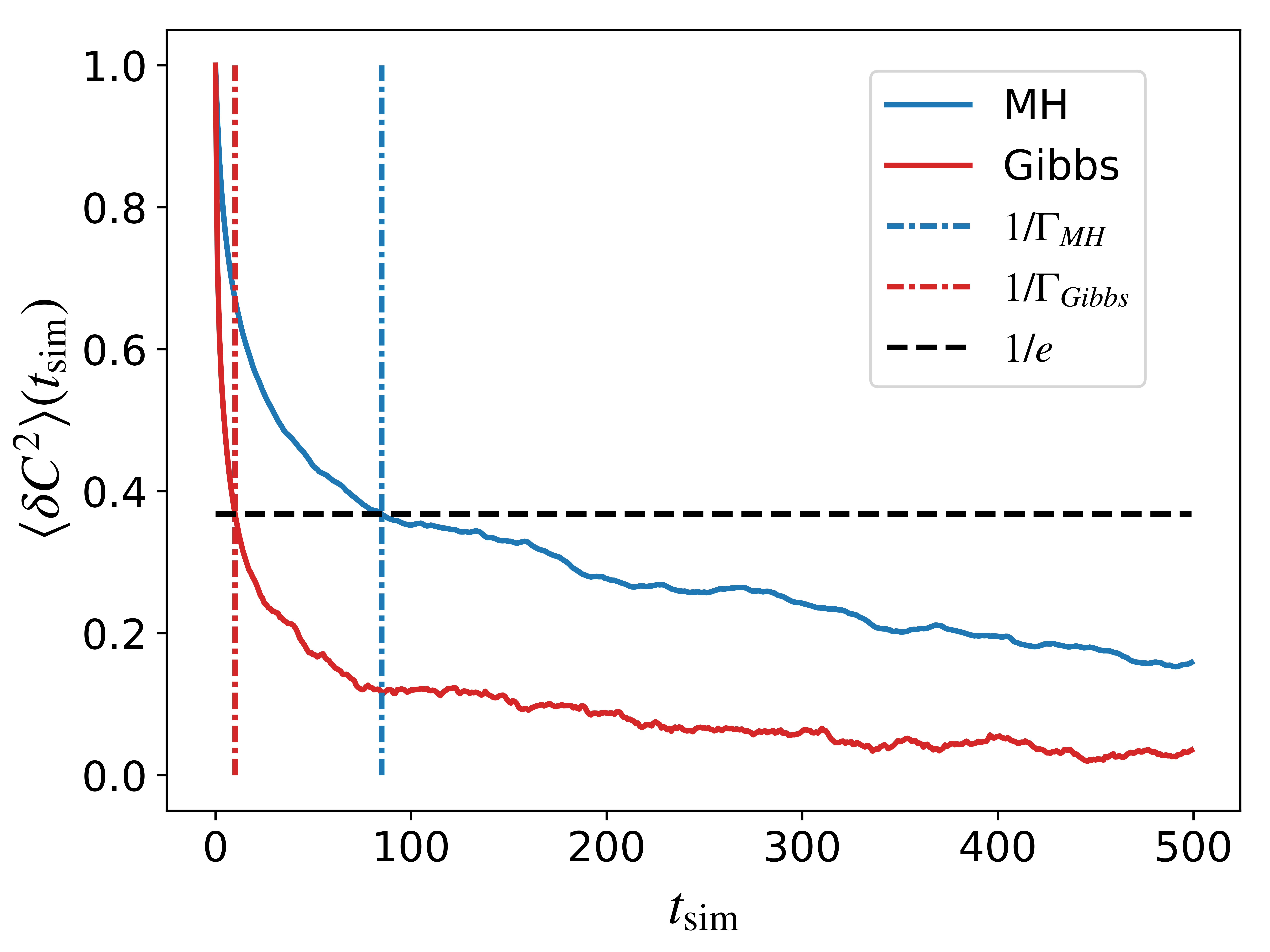}
    \includegraphics[width=.33\textwidth]{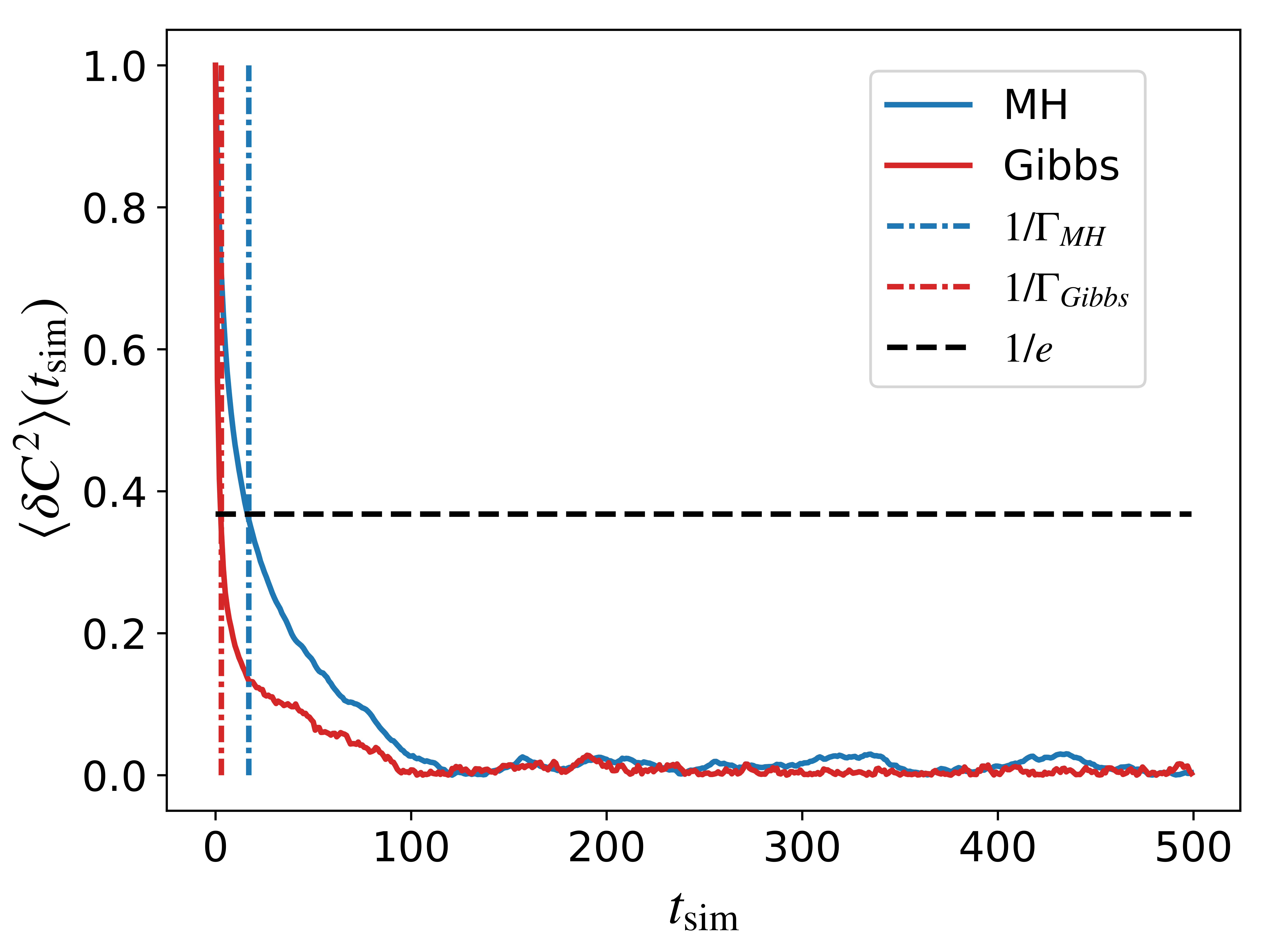}
    \caption{Markov chain autocorrelation for the M-H and  Gibbs sampling algorithms computed using the orientational correlation for $g = 0.1$ (left panel), $g = 0.5$ (middle panel) and $g = 1.0$ (right panel). All three simulations used $N = 100$, $P = 48$ and $50000$ simulation steps.} \label{autocorr}
    \end{figure*}
    
\subsection{Correlation analysis}
As mentioned above, the Gibbs sampling algorithm yields faster Markov chain decorrelation. Consequently, less Monte Carlo steps are required for a given simulation. To exhibit this, we computed the chain autocorrelation for the M-H algorithm as well as the Gibbs sampling algorithm (see Fig. \ref{autocorr}) using the orientational correlation as the estimator for $g = 0.1$, $g = 0.5$ and $g = 1.0$. All three simulations used $N = 100$, $P = 48$ and $50000$ simulation steps. For $g = 0.1$, the starting configuration is random for each rotor whereas for $g = 0.5$ and $g = 1.0$, the simulation is initiated with all the rotors aligned (pointing along the chain). Then, we computed the decorrelation time $1/\Gamma$ for each simulation: 
\begin{equation}
    1/\Gamma = \text{inf}\{t \ | \ \langle \delta C^{2} \rangle < 1/e\}
\end{equation}
where $t$ here corresponds to simulation steps, $\langle \delta C^{2} \rangle $ is the autocorrelation of the orientational correlation estimator and $\text{inf}$ denotes the infimum of the set in question. In all cases, the Gibbs decorrelation time is less than the M-H decorrelation time which implies that the Gibbs sampling algorithm outperforms the M-H algorithm for this system for all phases (ordered as well as disordered).


In the critical region ($g = 0.5$), the decorrelation time for the M-H algorithm is much larger than in the ordered and disordered phases as shown in Fig. \ref{autocorr}. This is characteristic of critical slowing down near phase transitions. However, it is interesting to note that the Gibbs algorithm autocorrelation decays very fast here as well which allows us to accurately estimate the orientational correlation in the critical region as shown in Fig. \ref{PIGS_DMRG_corr}. 

\section{Concluding Remarks\label{conclusions}}
We developed a path integral Monte Carlo simulation approach based on the discretization of continuous degrees of freedom and Gibbs sampling.  The method is general and tested on a many-body problem, namely the dipolar planar rotor chain.
The DVR is used as a discretization scheme and allows the representation of individual kinetic energy contributions as $1$-body off-diagonal density matrices. The interaction potential is diagonal in the DVR which leads to a great simplification of its contribution to the path integral action. One key advantage of the DVR is that the path integral becomes a path sum and each state on the multidimensional grid has a finite probability. One can therefore sample the configuration space of the path integral by constructing conditional probability tables and systematically sample all degrees of freedom without rejection using Gibbs sampling.
The approach was initially tested for small chains with $N\leq3$ where ED and NMM calculations are feasible. These calculations also allow one to determine the critical $\tau$ value above which the $1$-body kinetic density matrix remains positive definite. We suggest performing such a test before embarking in many-body simulations with large $N$ values. One can then extrapolate to $\tau \rightarrow 0$ by performing a series of simulations with different $\tau$ values (numbers of beads) for a given $\beta$.
The  method is tested for a large many-body system consisting of $N=100$ interacting rotors. Grounds state energies and correlation functions are in excellent agreement with benchmark DMRG calculations for a broad range of $g$ values.
This system size is sufficiently large to exhibit a phase transition at a critical value of $g_c\sim 0.5$. Both the disordered phase and the ordered phase are accurately captured by our method. We observe that the use of Gibbs sampling allows for shorter decorrelation times than usual rejection based M-H sampling. This makes the new sampling scheme very advantageous for the simulation of complex materials. Indeed, the approach will be used to study the nature of water molecules trapped in beryl\cite{gorshunov2016incipient,kolesnikov2016quantum,zhukova2019quantum,belyanchikov2020dielectric} 
and cordierite\cite{belyanchikov2022single} in future work.

\section*{Acknowledgements}
We thank Adrian Del Maestro and Juan Carrasguilla for useful discussions.
The authors acknowledge the Natural Sciences and Engineering Research Council (NSERC) of Canada, the Ontario Ministry of Research and Innovation (MRI), the Canada Research Chair program, and the Canada Foundation for Innovation (CFI).

\section*{DATA AVAILABILITY}
The data that support the findings of this study are available from the corresponding author upon reasonable request.

\bibliography{literature}

\end{document}


\section{Determination of number of grids}
\section{Test convergence of different interaction strength g}
\subsection{tau convergence}
\begin{figure}[htbp]
    \centering
    \begin{subfigure}[b]{0.48\textwidth}
        \includegraphics[width=\textwidth]{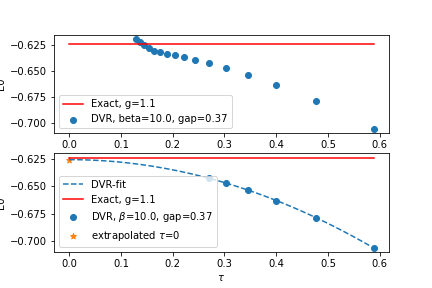}
        \caption{2 rotors}
        \label{fig:sub1}
    \end{subfigure}
    \hfill 
    \begin{subfigure}[b]{0.48\textwidth}
        \includegraphics[width=\textwidth]{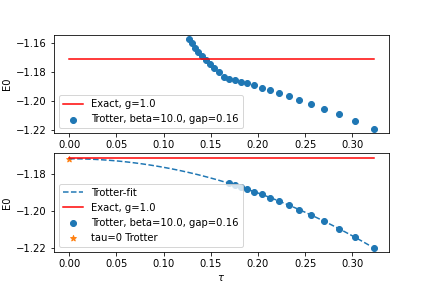}
        \caption{3 rotors}
        \label{fig:sub2}
    \end{subfigure}
    \caption{tau convergence}
    \label{fig:test}
\end{figure}

\subsection{beta convergence and test}

\begin{figure}[htbp]
    \centering
    \begin{subfigure}[b]{0.48\textwidth}
        \includegraphics[width=\textwidth]{E0_vs_beta_N2tau0.2g1.1(bestP7).png}
        \caption{beta convergence}
        \label{fig:sub1}
    \end{subfigure}
    \hfill 
    \begin{subfigure}[b]{0.48\textwidth}
        \includegraphics[width=\textwidth]{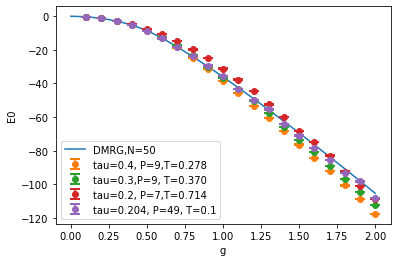}
        \caption{test with groundstate energy}
        \label{fig:sub2}
    \end{subfigure}
    \caption{beta convergence}
    \label{fig:test}
\end{figure}

\section{bench marking with DMRG}

\begin{figure}[htbp]
    \centering
    \begin{subfigure}[b]{0.48\textwidth}
        \includegraphics[width=\textwidth]{E0_vs_g_N10_tau0.186.png}
        \caption{2 rotors}
        \label{fig:sub1}
    \end{subfigure}
    \hfill 
    \begin{subfigure}[b]{0.48\textwidth}
        \includegraphics[width=\textwidth]{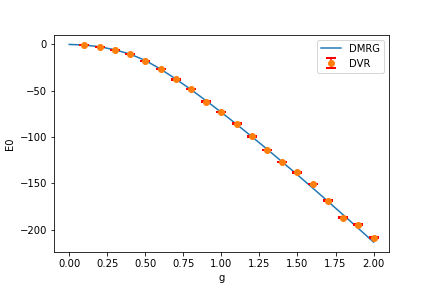}
        \caption{4 rotors}
        \label{fig:sub2}
    \end{subfigure}
    \caption{Benchmarking with DMRG-1}
    \label{fig:test}
\end{figure}

\begin{figure}[htbp]
    \centering
    \begin{subfigure}[b]{0.48\textwidth}
        \includegraphics[width=\textwidth]{E0_vs_g_N10_tau0.186.png}
        \caption{6 rotors}
        \label{fig:sub1}
    \end{subfigure}
    \hfill 
    \begin{subfigure}[b]{0.48\textwidth}
        \includegraphics[width=\textwidth]{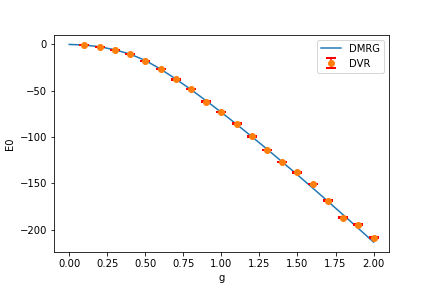}
        \caption{8 rotors}
        \label{fig:sub2}
    \end{subfigure}
    \caption{Benchmarking with DMRG-2}
    \label{fig:test}
\end{figure}

\begin{figure}[htbp]
    \centering
    \begin{subfigure}[b]{0.48\textwidth}
        \includegraphics[width=\textwidth]{E0_vs_g_N10_tau0.186.png}
        \caption{10 rotors}
        \label{fig:sub1}
    \end{subfigure}
    \hfill 
    \begin{subfigure}[b]{0.48\textwidth}
        \includegraphics[width=\textwidth]{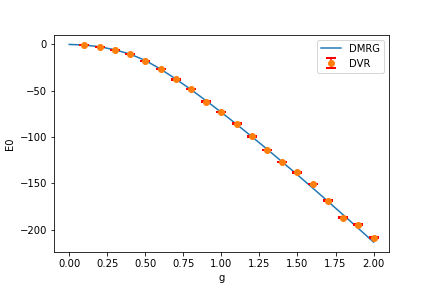}
        \caption{100 rotors}
        \label{fig:sub2}
    \end{subfigure}
    \caption{Benchmarking with DMRG-3}
    \label{fig:test}
\end{figure}

\section{Rotor orientation analysis}
\subsection{initial PIMC}
\begin{figure}
    \centering
    \includegraphics[width=0.9\columnwidth]{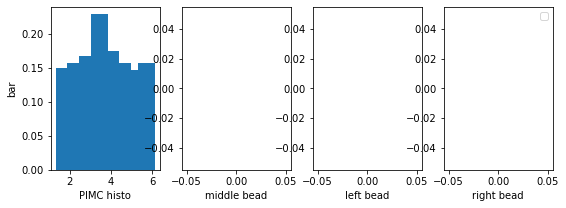}
    \caption{PIMC initial histograph $N=10$ $g=0.1$}
    \label{fig:enter-label}
\end{figure}

\begin{figure}
    \centering
    \includegraphics[width=0.9\columnwidth]{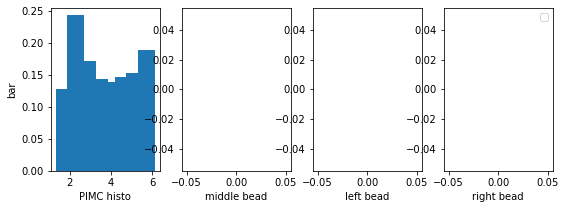}
    \caption{PIMC initial histograph $N=10$ $g=0.5$}
    \label{fig:enter-label}
\end{figure}

\begin{figure}
    \centering
    \includegraphics[width=0.9\columnwidth]{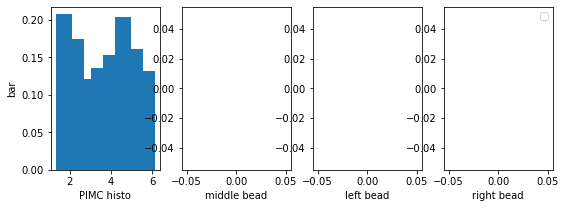}
    \caption{PIMC initial histograph $N=10$ $g=1.0$}
    \label{fig:enter-label}
\end{figure}

\subsection{after PIMC}

\subsubsection{terminal rotor}

\begin{figure}
    \centering
    \includegraphics[width=0.9\columnwidth]{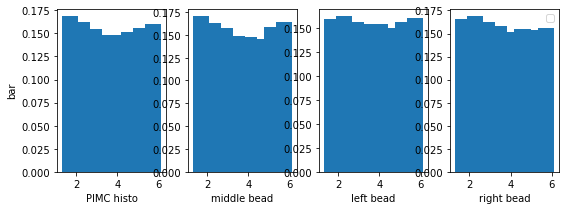}
    \caption{Terminao rotor orientation histograph $N=10$ $g=0.1$}
    \label{fig:enter-label}
\end{figure}

\begin{figure}
    \centering
    \includegraphics[width=0.9\columnwidth]{TerminalRotor_g0.5.png}
    \caption{Terminao rotor orientation  histograph $N=10$ $g=0.5$}
    \label{fig:enter-label}
\end{figure}

\begin{figure}
    \centering[H]
    \includegraphics[width=0.9\columnwidth]{TerminalRotor_g1.0.png}
    \caption{Terminao rotor orientation  histograph $N=10$ $g=1.0$}
    \label{fig:enter-label}
\end{figure}

\begin{figure}[h]
    \centering
    \includegraphics[width=0.9\columnwidth]{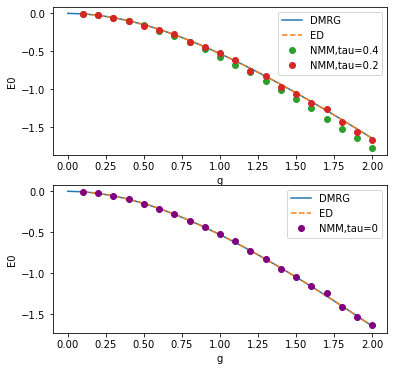}
    \caption{Moved to supplementary info from main text - figure 4 in main text has the energy result for N = 2 and 3}
    \label{fig:benchmarking}
\end{figure}

\subsection{after PIMC}